\documentclass[11pt]{article}
\begin{document}
\title
{No hair theorems for stationary axisymmetric black holes}
\author
{Sourav Bhattacharya{\footnote{sbhatt@bose.res.in }}
 \quad and Amitabha Lahiri{\footnote{amitabha@bose.res.in}}\\
S. N. Bose National Centre for Basic Sciences, \\
JD Block, Sector III, Salt Lake, Kolkata -700098, India.\\
}

\maketitle
\abstract

{We present a non-perturbative proof of the no hair theorems
  corresponding to scalar and Proca fields for stationary
  axisymmetric de Sitter black hole spacetimes. Our method also
  applies to asymptotically flat and under a reasonable assumption,
  to asymptotically anti-de Sitter spacetimes.}

\hskip 1cm

    {\bf Keywords:} {Stationary black holes,
 no hair theorem, de Sitter 

\vskip 2cm

\section{Introduction}
The classical no hair conjecture for black holes states that any
gravitational collapse reaches a final stationary state
characterized only by a small number of parameters. A part of this
conjecture has been proven rigorously by taking different matter
fields, known as the no hair theorem~(see
e.g. \cite{Chrusciel:1994sn, Heusler:1998ua,Bekenstein:1998aw}) and
deals with the uniqueness of stationary black hole solutions
characterized only by mass, angular momentum, and charges
corresponding to long range gauge fields such as the
electromagnetic field. Any non-trivial field configuration other
than the long range gauge fields present at the exterior of a
stationary black hole is known as `hair'.  In particular, it has
been shown that static, spherically symmetric black holes do not
support hair corresponding to scalars in convex potentials,
Proca-massive vector fields~\cite{Bekenstein:1971hc}, or even gauge
fields corresponding to the Abelian Higgs model~\cite{Adler:1978dp,
  Lahiri:1993vg}.

All the above proofs assume the spacetime to be asymptotically
flat, i.e., one can reach spacelike infinity so that sufficiently
rapid fall-off conditions on the matter fields can be imposed
there. But recent observations~\cite{Riess:1998cb,
  Perlmutter:1998np} suggest that there is a strong possibility
that our universe is endowed with a small but positive cosmological
constant $\Lambda$. It is generally expected that in that case the
spacetime in its stationary state should have an outer or
cosmological Killing horizon~\cite{Bhattacharya:2010vr}. The
cosmological Killing horizon acts in general as a causal boundary
(see e.g. \cite{Gibbons:1977mu}) so that no physical observer can
communicate beyond this horizon along a future directed path.  If
there is a black hole, the black hole event horizon will be located
inside the cosmological horizon and the spacetime is then known as
a de Sitter black hole spacetime. The observed value of the
$\Lambda$ is very small, of the order of $ 10^{-52} {\rm{m}}^{-2}$,
and the known exact solutions~\cite{Carter:1968ks} for a small
$\Lambda$ suggest that the cosmological horizon has a length scale
$\sim {\cal{O}}\left(\frac{1}{\sqrt{\Lambda}}\right)$ which is of
course large, but not infinite. Since no physical observer can
communicate beyond the cosmological horizon, in a de Sitter black
hole spacetime the cosmological horizn acts as a natural
boundary. So in the most general case one cannot assume any precise
asymptotic form in the vicinity of the cosmological horizon and
hence one cannot set $T_{ab}=0$ over that horizon. Therefore, the
extension of the no hair theorems for de Sitter black holes are
expected to be different from the $\Lambda \leq 0$ cases.

In particular, a lot of progress has been made in this topic for
static de Sitter black holes. Price's theorem, which can be
regarded as a perturbative no hair theorem~\cite{Price:1971gc} was
proved in~ \cite{Chambers:1994sz} for a Schwarzschild-de Sitter
background by taking massless perturbations.
In~\cite{Bhattacharya:2007ap}, all the known black hole no hair
theorems were extended for a general static de Sitter black hole
spacetime. The exception was that a charged solution corresponding
to the false vacuum of the complex scalar of the Abelian Higgs
model was obtained which has no $\Lambda\leq 0$ analogue. In fact
this charged solution suggests that even though $\Lambda$ is very
small, the existence of the cosmological horizon, because of the
non-trivial boundary conditions, may change local physics
considerably.

It is thus an interesting task to generalize the no hair theorems
for a stationary de Sitter black hole. For an asymptotically flat
spacetime, the no hair proofs for a rotating black hole for scalar
and Proca fields were first given in~\cite{Bekenstein:1972ky}
assuming time reversal symmetry of the matter equations. For a
discussion on the 2+1 dimensional no hair theorem
see~\cite{Skakala:2009ss}. See also~\cite{Sen:1998bj} for a scalar
no hair theorem in stationary asymptotically flat spacetimes with
non-minimally coupled scalar fields.  In the following we shall
give a proof of the no hair theorems for scalar and Proca massive
vector fields for a de Sitter black hole spacetime. Our method will
be considerably different from that of~\cite{Bekenstein:1972ky}.

This paper is organized as follows. In the next section we outline
all the necessary assumptions and the geometrical set up we work
in. In Sec. 3 we give the proof of the no hair theorems for the
scalar and Proca fields. Finally we discuss our results. We set
$c=1=G$ throughout.

\section{The geometrical set up}
In this section we outline the particular geometrical set up we
need to describe our spacetime. More details can be found in
\cite{Bhattacharya:2010vr}.

We consider a $(3+1)$-dimensional stationary axisymmetric spacetime
with two commuting Killing fields $\left\{\xi^a,~\phi^a\right\}$,
\begin{eqnarray}
\nabla_{(a}\xi_{b)} &=& 0 =
 \nabla_{(a}\phi_{b)} \,,\\
\left[\xi, \phi\right]^a &=& 0\,.
\label{g1}
\end{eqnarray}
$\xi^a$ is locally timelike with norm $\xi^a\xi_a = -\lambda^2$ and
generates the stationarity, whereas $\phi^a$ is locally spacelike
with closed orbits and norm $\phi^a\phi_a = f^2$ and generates the
axisymmetry. We assume that the spacetime satisfies Einstein's
equations.  We take the connection $\nabla_a$ to be torsion free,
i.e., for any differentiable function $g(X)$ we have
$\nabla_{[a}\nabla_{b]} g(X)=0$.

We can specify a basis $\left\{\xi^a,~\phi^a,~\mu^a,~\nu^a\right\}$
for this spacetime, where $\left\{\mu^a,~\nu^a\right\}$ are
spacelike basis vectors orthogonal to both $\xi^a$ and $\phi^a$. We
assume that the 2-surfaces spanned by $\left\{\mu^a,~\nu^a\right\}$
form integral submanifolds. In other words,
$\left\{\mu^a,~\nu^a\right\}$ form the basis of a Lie algebra. We
note that this assumption is valid for known stationary
axisymmetric spacetimes.

A stationary axisymmetric spacetime with a black hole is in general
rotating. In that case $\xi^a$ is not orthogonal to $\phi^a$, and
the basis $\left\{\xi^a,~\phi^a,~\mu^a,~\nu^a\right\}$ is not
orthogonal. So in particular, there is no family of spacelike
hypersurfaces which is both tangent to $\phi^a$ and orthogonal to
$\xi^a$. Let us first construct a family of spacelike hypersurfaces
tangent to $\phi^a$. We define $\chi_{a}$ as
\begin{eqnarray}
\chi_a=\xi_a-\frac{1}{f^2}\left(\xi_b\phi^b\right)
\phi_a \equiv \xi_a+\alpha \phi_a,
\label{g2}
\end{eqnarray}
so that we have $\chi_a\phi^a=0$ everywhere. We note that
\begin{eqnarray}
\chi_a\chi^a  = -\left(\lambda^2+\alpha^2 f^2\right),
\label{g3}
\end{eqnarray}
so that $\chi_a$ is timelike when $\beta^2=
\left(\lambda^2+\alpha^2 f^2\right)>0$. The basis
$\left\{\chi^a,~\phi^a,~\mu^a,~\nu^a\right\}$ is now an orthogonal
basis for the spacetime.  We also have
\begin{eqnarray}
\nabla_{(a}\chi_{b)} = \phi_a\nabla_b \alpha
+\phi_b\nabla_a \alpha.
\label{g4}
\end{eqnarray}
Our assumption that $\{\mu^a,~\nu^a\}$ span an integral
2-submanifold implies that $\chi^a$ satisfies the Frobenius
condition of hypersurface orthogonality \cite{Bhattacharya:2010vr}
\begin{eqnarray}
\chi_{[a}\nabla_b\chi_{c]}=0.
\label{g5}
\end{eqnarray}
Thus $\chi^a$ is orthogonal to the spacelike
$\left\{\phi^a,~\mu^a,~\nu^a\right\}$ hypersurfaces, say $\Sigma$.

How do we define the horizons of our spacetime? It is known that in
a rotating black hole spacetime, $\xi^a$ becomes spacelike within
the ergosphere~\cite{Wald:1984rg}, so for such spacetimes
$\lambda^2=0$ does not in general define a horizon.  It was shown
in \cite{Bhattacharya:2010vr} by considering the null geodesic
congruence over a $\beta^2=0$ surface that the vector field
$\chi^a$ coincides with a null Killing field over that
surface. Thus a $\beta^2=0$ surface is essentially a Killing or
true horizon. Accordingly, we define the black hole event horizon
and the cosmological event horizon to be the two $\beta^2=0$
surfaces. An example of this is the Kerr-Newman-de Sitter spacetime
\cite{Gibbons:1977mu}.

We assume that no naked curvature singularity exists anywhere in
our region of interest, i.e., anywhere between the two
horizons. The Einstein equation $G_{ab}+\Lambda g_{ab}=T_{ab}$ then
implies that the invariants constructed from the energy-momentum
tensor $T_{ab}$ are bounded over or everywhere in the region
between the two horizons. Apart from this regularity, we also
assume that the horizons are `closed' surfaces.

The usual projector or the induced metric over the spacelike
hypersurfaces $\Sigma$ is defined as
\begin{eqnarray}
h_{a}{}^{b}=\delta_{a}{}^{b}+\beta^{-2}\chi_a\chi^b.
\label{g6}
\end{eqnarray}
Let $D_a$ be the induced connection over $\Sigma$ defined via the
projector as $D_a:=h_{a}{}^{b}\nabla_b$. Then we can project the
derivative of a tensor $T_{a_1a_2\cdots}{}^{b_1b_2\cdots}$ over
$\Sigma$ as
\begin{eqnarray}
D_a\widetilde{T}_{a_1a_2\dots}{}^{b_1b_2\dots}:=h_{a}{}^{b}
 h_{a_1}{}^{c_1}\dots h^{b_1}{}_{d_1}\dots\nabla_b
T_{c_1c_2\dots}{}^{d_1d_2\dots},
\label{g7}
\end{eqnarray}
where $\widetilde{T}$ is the projection of $T$ over $\Sigma,$ given
by $\widetilde{T}_{a_1a_2\cdots}{}^{b_1b_2\cdots} :=
h_{a_1}{}^{c_1}\cdots h^{b_1}{}_{d_1}\cdots
T_{c_1c_2\cdots}{}^{d_1d_2\cdots}$. It is easy to verify that the
induced connection $D_a$ over $\Sigma$ defined in Eq.~(\ref{g7})
satisfies the Leibniz rule and is compatible with the induced
metric $h_{ab}$.

It will be useful to note here that if a function $\psi$ has a
vanishing Lie derivative with respect to $\chi$, that is if
$\pounds_{\chi}\psi=0$, we can use the torsion free condition to
write 
\begin{eqnarray}
\beta\nabla_a\nabla^a \psi=D_a\left(\beta D^a\psi\right).
\label{g8}
\end{eqnarray}
Next we note that the subspace spanned by
$\left\{\chi^a,~\mu^a,~\nu^a \right\}$ do not form a hypersurface.
This is because the necessary and sufficient condition that an
arbitrary subspace of a manifold forms an integral submanifold or a
hypersurface is that the basis vectors of that subspace span a Lie
algebra (see e.g.~\cite{Wald:1984rg} and references therein). It is
easy to verify using the definition of $\chi^a$ in Eq.~(\ref{g2})
that the basis vectors $\left\{\chi^a,~\mu^a,~\nu^a\right\}$ do not
span a Lie algebra. This implies that we cannot write a condition
like $\phi_{[a}\nabla_b\phi_{c]}=0$ \cite{Bhattacharya:2010vr}.

However, according to our assumptions, there are integral spacelike
2-submanifolds orthogonal to both $\chi^a$ and $\phi^a$, and
spanned by $\left\{\mu^a,~\nu^a\right\}$. Then over
these 2-manifolds $\overline{\Sigma},$  we must have 
\begin{eqnarray}
\phi_{[a}D_{b}\phi_{c]}=0.
\label{g9}
\end{eqnarray}
Using the projector defined in Eq.~(\ref{g6}), we write the Killing
equation for $\phi_a$ over $\Sigma$ as
\begin{eqnarray}
D_{(a}\phi_{b)}=0.
\label{add1}
\end{eqnarray}
We now solve Eq.s (\ref{g9}) and (\ref{add1}) to find the
expression for $D_a\phi_b$ and using the projector (\ref{g6})
rewrite it in terms of the full spacetime connection $\nabla_a$
\begin{eqnarray}
\nabla_a\phi_b=\frac{1}{f}
\left[\phi_b \nabla_a f-\phi_a \nabla_b f\right]+
\frac{f^2}{2\beta^2}
\left[\chi_b \nabla_a \alpha
-\chi_a \nabla_b \alpha \right].
\label{g10}
\end{eqnarray}
Also, we note that since $\left\{\mu^a,~\nu^a\right\}$ span
integral 2-surfaces $\overline{\Sigma}$, and $\chi^a$ and $\phi^a$
are orthogonal, we can project spacetime tensors over
$\overline{\Sigma}$ via the projector
\begin{eqnarray}
\Pi_{a}{}^{b}=\delta_{a}{}^{b} +
\beta^{-2}\chi_a\chi^b-f^{-2}\phi_a\phi^b.   
\label{g11}
\end{eqnarray}
We can also define the induced connection $\overline{D}_a$ on
$\overline{\Sigma}$ using the projector $\Pi_{a}{}^{b}$. 

We will assume that any matter field also obeys the symmetry of the
spacetime. In other words, if $X$ is a matter field, or a component
of a matter field, we must have
\begin{eqnarray}
\pounds_{\xi}X=0=\pounds_{\phi}X.
\label{sym}
\end{eqnarray}
We note that Eq.~(\ref{sym}) need not hold if $X$ is a gauge field.
 
We are now ready to prove the no-hair theorems.

\section{No hair theorems for scalar and Proca fields} 
We start with the simplest case, that of a scalar field $\psi$
moving in a potential $V(\psi)$ satisfying the equation of motion
\begin{eqnarray}
\nabla_a\nabla^a\psi= V'(\psi),
\label{nh1}
\end{eqnarray}
where the `prime' denotes differentiation with respect to $\psi$
and any mass term is included in $V(\psi)$.  Since we are assuming
stationarity and axisymmetry, we must have $\pounds_{\xi}\psi=0 =
\pounds_{\phi}\psi$, as we mentioned earlier. Since $\chi_a=\xi_a +
\alpha\phi_a$, it follows that $\pounds_{\chi}\psi=0$. Then using
Eq.~(\ref{g8}) we find that Eq.~(\ref{nh1}) takes the following
form over the $\chi$-orthogonal hypersurface $\Sigma$,
\begin{eqnarray}
D_a\left(\beta D^a\psi\right)=\beta V'(\psi).
\label{nh2}
\end{eqnarray}
We now multiply Eq.~(\ref{nh2}) by $V'(\psi)$ and integrate by
parts to have 
\begin{eqnarray}
\int_{\partial \Sigma}\beta
V'(\psi)n^aD_a\psi 
+\int_{\Sigma}\beta\left[V''(\psi) 
\left(D^a\psi\right)\left(D_a\psi\right)
+ V'^2(\psi)\right]=0,
\label{nh3}
\end{eqnarray}
where $\partial\Sigma$ are spacelike closed 2-surfaces located at
the boundaries of $\Sigma$, i.e., the horizons and $n^a$ is a unit
spacelike vector normal to these 2-surfaces.

According to our assumption, there is no naked curvature
singularity anywhere between the horizons, including the
horizons. This implies that the invariants of the energy momentum
tensor is bounded on the horizons. Since $\nabla_a\psi\nabla^a\psi$
appears in the trace of the energy-momentum tensor, it follows that
this quantity is bounded on the horizons. On the other hand,
$\pounds_{\chi}\psi=0$ implies that $\nabla_a\psi=D_a\psi$, while
the inequality $\left(D_a\psi-n_a
  \left(n^bD_b\psi\right)\right)^2\geq0$ implies $\big\vert
n^aD_a\psi\big\vert^2 \leq \left(D_a\psi\right)\left(D^a
  \psi\right)$. Therefore the quantity $n^aD_a\psi$ also remains
bounded over the horizons. Then since $\beta=0$ over the horizons,
the surface integrals in Eq.~(\ref{nh3}) vanish.

Since the inner product in the $\Sigma$ integral of Eq.~(\ref{nh3})
is spacelike, it immediately follows that no non-trivial solution
exists for $\psi$ over $\Sigma$ for a convex potential, i.e., if
$V''(\psi)>0$ for all values of $\psi$. So for a convex $V(\psi)$
the scalar field $\psi$ is a constant located at the minimum of the
potential $V(\psi)$.  Then $\pounds_{\chi}\psi=0$ ensures that we
have the same trivial solution throughout the spacetime. This is
the standard no hair result for a scalar field.

For $V(\psi)=0$, we multiply Eq. (\ref{nh2}) by $\psi$ and
integrate by parts over $\Sigma$ to get an equation similar to
Eq.~(\ref{nh3}). Assuming that $\psi$ is measurable, i.e. bounded,
over the horizon ~\cite{ Bekenstein:1971hc,
Bekenstein:1972ky}, gives the no hair result.

The no-hair statement need not hold in other kinds of
potentials. For static spherically symmetric spacetimes scalar hair
may be present for non-convex potentials, such as the double well
potential $V(\psi)=\frac{\lambda}{4}(\psi^2-v^2)^2$, which gives an
unstable solution \cite{Torii:1998ir}. Another example is that of a
conformal scalar $\psi$ coupled to gravity by a term $V(\psi) =
\frac{1}{12}R\psi^2$. The scalar field action is invariant under a
conformal transformation in this theory.  So by appropriately
choosing the conformal factor of the transformation we can make
make $\psi$ or $n^aD_a\psi$ diverge at $\partial\Sigma$ without
causing a curvature singularity. Then the $\partial\Sigma$ integral
can be non-zero, which allows a non-trivial configuration of $\psi$
on $\Sigma$. In fact static spherically symmetric solutions with
conformal scalar hair with $\Lambda>0$ are
known~\cite{Martinez:2002ru}. It is likely that these exceptions
will also be present for stationary axisymmetric spacetimes.

Next we consider the Proca massive Lagrangian for the vector field
\begin{eqnarray}
\mathcal{L} = -\frac{1}{4} F_{ab}F^{ab} - \frac{1}{2} m^2
A_{b}A^{b}, 
\label{nh4}
\end{eqnarray}
where $F_{ab}=\nabla_a A_b-\nabla_b A_a$. We shall see below that
proving a no-hair statement in this case is quite a bit more
complicated than in the case of a scalar field. The equation of
motion for $A^b$ is
\begin{eqnarray}
\nabla_aF^{ab} - m^2A^b = 0.
\label{nh5}
\end{eqnarray}
The procedure, as for the scalar field, will be to construct a
positive definite quadratic with a vanishing integral on $\Sigma$.
Let us start by defining the potential $\psi$ and the `electric'
field $e^a$ 
\begin{eqnarray}
\psi =\beta^{-1}\chi_a A^a;\qquad e^a = \beta^{-1}\chi_bF^{ab}.
\label{nh6}
\end{eqnarray}
The vanishing of the Lie derivatives of $\psi$
and $e_a$ along the Killing fields $\xi^a$ and $\phi^a$ imply
\begin{eqnarray}
\pounds_{\chi}\psi=0;\qquad \pounds_{\chi}e^a=-\phi^a e^b\nabla_b
\alpha. 
\label{nh7}
\end{eqnarray}
Then using Eq.s (\ref{g5}), (\ref{g6}), (\ref{g7}) it is easy to
obtain the following projected equations over $\Sigma$
\begin{eqnarray}
D_a(\beta \psi)=\beta e_a +\pounds_{\chi}A_a;\qquad D_ae^a=m^2\psi.
\label{nh7.1}
\end{eqnarray}
We
now multiply the second of the Eq.s (\ref{nh7.1}) with $\beta\psi$,
use the first of the Eq.s (\ref{nh7.1}) and integrate by parts over
$\Sigma$ to get
\begin{eqnarray}
\int_{\partial \Sigma}\beta \psi n^a e_a 
+\int_{\Sigma}\left[\beta \left(e_ae^a+m^2\psi^2\right) +e^a
  \left(\pounds_{\chi}A_a\right)\right]=0. 
\label{nh8}
\end{eqnarray}
Using the fact that $\pounds_{\xi}A_a=0=\pounds_{\phi}A_a$, we have
$\pounds_{\chi}A_a=\left(A_b\phi^b\right)\nabla_a \alpha$.  The
terms $\psi^2$ and $e_a^2$ appear in the invariants of the
energy-momentum tensor which are bounded over the horizons. This
implies that the surface integrals vanish, giving us the following
$\Sigma$ integral
\begin{eqnarray}
  \int_{\Sigma}\left[\beta \left(e_ae^a+ m^2\psi^2\right)
    + \left(A_b\phi^b\right)e^a\nabla_a \alpha \right]=0. 
\label{nh9}
\end{eqnarray}
We note that for $m=0$ the Lagrangian (\ref{nh4}) is invariant
under a local gauge symmetry $A\to A+dg$, where $g$ is any
differentiable function. Then for $m = 0$, the components of $A$
are not physical and need not be bounded on the horizon. Then
we can always choose 
$\psi$ such that the surface integrand in Eq. (\ref{nh8}) becomes
unbounded and hence the surface integral becomes non-zero.

By Eq.~(\ref{nh6}), $e^a\chi_a=0$ and hence $e^a$ is a spacelike
vector field. Also, $\beta >0$ between the two horizons and
vanishes on the horizons. So all but the last term in
Eq.~(\ref{nh9}) are positive definite. Nor can we set the last term
to zero, since $\chi^a$ is not a Killing field. Thus the no hair
conjecture for the Proca field cannot be proven from
Eq.~(\ref{nh9}) alone, and we need to take a more careful look at
the rest of the equations of motion.

Let us first project Eq.~(\ref{nh5}) over $\Sigma$. Let $a_b$ and
$f_{ab}$ be the $\Sigma$ projections of $A_b$ and $F_{ab}$ defined
via the projector as $a_b:=h_{b}{}^{a}A_a;~f_{ab}:=
h_{a}{}^{c}h_{b}{}^{d}F_{cd}$. It is easy to see that
\begin{eqnarray}
h_{a}{}^{c}h_{b}{}^{d}F_{cd}=D_a a_b-D_b a_a.
\label{nh10}
\end{eqnarray}
We now multiply Eq.~(\ref{nh5}) by the projector to write
\begin{eqnarray}
\beta h^{b}{}_{c}\nabla_a F^{ac}=m^2\beta a^b.
\label{nh11}
\end{eqnarray}
To relate Eq.~(\ref{nh11}) to the induced connection $D_a$ and the
projected tensor $f_{ab}$ we consider the expression
$D_{a}\left(\beta f^{ab}\right)$. Using the definition of the
projector, we can write
\begin{eqnarray}
D_{a}\left(\beta f^{ab}\right)&=& h^{b}{}_{e}h^{f}{}_{a}\nabla_f
\left(\beta F^{ae}\right) \nonumber \\
&=& h^{b}{}_{e} \nabla_a\left(\beta
  F^{ae}\right)+\beta^{-2}h^{b}{}_{e}\chi_a\chi^f\nabla_f
\left(\beta F^{ae}\right). 
\label{add2}
\end{eqnarray}
The orthogonality of $\chi_a$ and $\phi_a$ and Eq.~(\ref{g4}) imply 
$\pounds_{\chi}\beta=0$.  Also, since $\xi^a$ and $\phi^a$ are
Killing fields we have $\pounds
_{\xi}F^{ab}=0=\pounds_{\phi}F^{ab}$. Then Eq.~(\ref{add2}) becomes 
\begin{eqnarray}
D_{a}\left(\beta f^{ab}\right)=\beta h^{b}{}_{e} \nabla_a F^{ae}
+\beta^{-1}\chi_a h^{b}{}_{e}\left[F^{ce}\nabla_c
  \chi^a+F^{ac}\nabla_c\chi^e -\left(F^{ce}\nabla_c\alpha\right)
  \phi^a -\left(F^{ac}\nabla_c \alpha\right) \phi^e\right]
\nonumber \\ 
 +h^{b}{}_{e}F^{ae}\nabla_a \beta.
\label{nh12}
\end{eqnarray}
On the other hand, from Eq.s (\ref{g4}) and (\ref{g5}) we have
\begin{eqnarray}
\nabla_a \chi_b=\beta^{-1}\left(\chi_b \nabla_a
  \beta-\chi_a\nabla_b \beta\right)
+\frac{1}{2}\left(\phi_a\nabla_b
  \alpha+\phi_b\nabla_a\alpha\right). 
\label{nh13}
\end{eqnarray}
We substitute this expression into Eq.~(\ref{nh12}). 
Then using $\chi^a\phi_a=0$ and the definition for the electric
field $e^a$, we find that Eq.~(\ref{nh12}) reduces to
\begin{eqnarray}
D_{a}\left(\beta f^{ab}\right)=\beta h^{b}{}_{e} \nabla_a
F^{ae}+\frac{1}{2} 
\left(e^c\nabla_c\alpha\right)\phi^b.
\label{nh14}
\end{eqnarray}
Thus Eq.~(\ref{nh11}) becomes
\begin{eqnarray}
D_{a}\left(\beta f^{ab}\right)=m^2\beta a^b+\frac{1}{2}
\left(e^c\nabla_c\alpha\right)\phi^b.
\label{nhn}
\end{eqnarray}
If we multiply both sides of Eq.~(\ref{nhn}) by $a_b$ and integrate
it over $\Sigma$, we again end up with an integral which, like
Eq.~(\ref{nh9}), is not guaranteed to be positive definite.

In order to simplify the situation, we now further project
Eq.~(\ref{nhn}) over the spacelike 2-submanifolds orthogonal to
both $\chi_a$ and $\phi_a$, which we have assumed to exist. We use
the projector $\Pi_{a}{}^{b}$ defined in Eq.~(\ref{g11}) and follow
the same procedure as before. Since $\phi^a$ is a Killing field,
$\pounds_{\phi}f_{ab}=0=\pounds_{\phi}a_b$ and we simply have after
a little computation
\begin{eqnarray}
\overline{D}_{a}\left(f\beta \overline{f}^{ab}\right)=m^2f\beta
\overline{a}^b, 
\label{nh14a}
\end{eqnarray}
where the `bar' denotes the respective fields after projection onto
these spacelike 2-submanifolds.  Contracting both sides of
Eq.~(\ref{nh14a}) by $\overline{a}_b$, integrating by parts and
using the same boundedness arguments over the horizons as before we
have
\begin{eqnarray}
\int_{\overline{\Sigma}}\beta f\left(\overline{f}_{ab}
  \overline{f}^{ab}+m^2\overline{a}^b\overline{a}_b\right)=0. 
\label{nh15}
\end{eqnarray}
Since the 2-submanifolds are spacelike the integrand in
Eq.~(\ref{nh15}) is positive definite. This yields
$\overline{f}_{ab}=0=\overline{a}_b$ everywhere over the
2-submanifolds. Also, it is easy to check using
$\pounds_{\xi}\overline{a}_b=0= \pounds_{\xi}\overline{f}_{ab}$ and
$\pounds_{\phi}\overline{a}_b=0=\pounds_{\phi}\overline{f}_{ab}$
that
$\pounds_{\chi}\overline{a}_b=0=\pounds_{\chi}\overline{f}_{ab}$.
This implies that $\overline{f}_{ab}=0=\overline{a}_b$ throughout
the manifold.

It follows that $A_b$ is of the form
\begin{eqnarray}
A_b=\Psi_1(x) \chi_b+\Psi_2(x)\phi_b.
\label{nh16}
\end{eqnarray}
The commutativity of the two Killing fields $\xi^a$ and $\phi^a$
implies that $\pounds_{\chi}\alpha=0=\pounds_{\phi}\alpha$. Also we
recall that since $A_a$ is a physical matter field, its Lie
derivatives vanish along $\xi^a$ and $\phi^a$. Then it is easy to
verify from Eq.~(\ref{nh16}) that $\pounds_{\phi}A_b=0$ implies
$\pounds_{\phi}\Psi_1=0=\pounds_{\phi}\Psi_2$; and
$\pounds_{\chi}A_b= \left(A_a\phi^a\right)\nabla_a \alpha$ implies
that $\pounds_{\chi}\Psi_1=0=\pounds_{\chi}\Psi_2$.

With the ansatz (\ref{nh16}),  the
Proca Lagrangian (\ref{nh4}) becomes
\begin{eqnarray}
  \mathcal{L}&=&\frac{1}{2}\left(\beta
    \nabla_a\Psi_1+2\Psi_1\nabla_a 
    \beta\right)^2-\frac{1}{2}\left(f\nabla_a\psi_2+2\Psi_2\nabla_a 
    f\right)^2+f^2\Psi_2\left(\nabla_a
    \psi_1\right)\left(\nabla^a\alpha\right) \nonumber \\ 
&& \qquad\qquad +\frac{f^4\Psi_2^2}{2\beta^2}\left(\nabla_a
  \alpha\right) 
  \left(\nabla^a\alpha\right) 
  +\frac{2f^2}{\beta}\Psi_1\Psi_2\left(\nabla_a\beta\right)
  \left(\nabla^a\alpha\right)+\frac{m^2}{2}\left(\beta^2  
    \Psi_1^2-f^2\Psi_2^2\right). 
\label{nh17}
\end{eqnarray}
The equations of motion for the two degrees of freedom $\Psi_1$ and
$\Psi_2$ are then
\begin{eqnarray}
 \nabla_a\left(\beta^2
  \nabla^a\Psi_1\right)-2\beta\left(\nabla_a\beta\right)
\left(\nabla^a \Psi_1\right)+\nabla_a\left(2\beta \psi_1\nabla^a
  \beta\right)-4\Psi_1\left(\nabla_a\beta\right)
\left(\nabla^a\beta\right) && \nonumber\\    +
\nabla_a\left(f^2\Psi_2\nabla^a\alpha\right)-
\frac{2f^2}{\beta}\Psi_2\left(\nabla_a\beta\right)
\left(\nabla^a\alpha\right)-m^2\beta^2\Psi_1&=&0, 
\label{nh18}
\end{eqnarray}
and
\begin{eqnarray}
\nabla_a\left(f^2 \nabla^a\Psi_2\right)-2f\left(\nabla_af\right)
\left(\nabla^a \Psi_2\right) + \nabla_a\left(2f
  \psi_2\nabla^af\right) -  
4\Psi_2\left(\nabla_af\right)\left(\nabla^af\right) 
\qquad\qquad &&\nonumber\\
+ \frac{f^4\Psi_2}{\beta^2}\left(\nabla_a\alpha
\right)\left(\nabla^a\alpha\right)
+ \frac{2f^2}{\beta}\Psi_1\left(\nabla_a\beta \right)
\left(\nabla^a\alpha\right)+f^2\left(\nabla_a\Psi_1\right)
\left(\nabla^a\alpha\right)-m^2f^2\Psi_2&=&0. 
\label{nh19}
\end{eqnarray}
Let us now project Eq.s (\ref{nh18}) and (\ref{nh19}) over $\Sigma$ 
and form quadratic integrals as before. Since
$\pounds_{\chi}\Psi_1=0=\pounds_{\chi}\Psi_2$, the fact that
$\nabla_a$ is torsion-free implies that
$\pounds_{\chi}\left(\nabla_a
  \Psi_1\right)=0=\pounds_{\chi}\left(\nabla_a \Psi_2\right)$.
It is straightforward to calculate similarly that $\pounds_{\chi}
\left(\nabla_a \alpha\right) = \pounds_{\chi}
\left(\nabla_af\right) = \pounds_{\chi} \left(\nabla_a \beta\right)
= 0.$ So the 1-forms $(\nabla_a \beta,~ \nabla_a
\alpha,~\nabla_af)$ are spacelike.  We can now project Eq.s
(\ref{nh18}) and (\ref{nh19}) over $\Sigma$ to get
\begin{eqnarray}
  D_a\left(\beta^3 D^a\Psi_1\right)-2\beta^2 \left(D_a\beta\right)
  \left(D^a \Psi_1\right)+D_a\left(2\beta^2 \Psi_1 D^a
    \beta\right)-4\beta\Psi_1 \left(D_a\beta
  \right)\left(D^a\beta\right)+\qquad \nonumber\\ 
  D_a\left(\beta f^2\Psi_2
    D^a\alpha\right)-
  2f^2\Psi_2\left(D_a\beta\right)
  \left(D^a\alpha\right)-m^2\beta^3\Psi_1=0, 
\label{nh20}
\end{eqnarray}
and
\begin{eqnarray}
D_a\left(f^2\beta D^a\Psi_2\right)-2\beta f\left(D_a f
\right)\left(D^a \Psi_2\right)+D_a\left(2\beta f
  \psi_2D^af\right)-4\beta \Psi_2\left(D_af\right)
\left(D^af\right)+\qquad\quad \nonumber\\ 
\frac{f^4\Psi_2}{\beta}\left(D_a\alpha \right)
\left(D^a\alpha\right)+
2f^2\Psi_1\left(D_a\beta\right)\left( D^a\alpha\right)+\beta
f^2\left(D_a\Psi_1\right)\left(D^a\alpha\right)-m^2\beta
f^2\Psi_2=0. 
\label{nh21}
\end{eqnarray}
We now multiply Eq.~(\ref{nh20}) by $\Psi_1$ and Eq.~(\ref{nh21})
by $\Psi_2$, add them and integrate by parts. The surface integrals
do not survive because $\Psi_1$ and $\Psi_2$ and their derivatives
are bounded on $\partial \Sigma,$ and we have
\begin{eqnarray}
\int_{\Sigma}\beta\left[ \left(\beta
    D_a\Psi_1+2\Psi_1D_a\beta\right)^2 + 
\left(fD_a\Psi_2+2\Psi_2D_a f\right)^2
-\frac{f^4\Psi_2^2}{\beta^2}\left(D_a\alpha\right)
\left(D^a\alpha\right) \right. \qquad\qquad \nonumber \\
\left. +m^2\left(\beta^2\Psi_1^2 +
  f^2\Psi_2^2\right)\right]=0. 
\label{nh22}
\end{eqnarray}
This is clearly not positive definite due to the presence of the
third term. We can naively interpret that term as the centrifugal
effect on the field due to the rotation of the spacetime. We now
investigate whether the rotation can actually be so large that the
integrand in Eq.~(\ref{nh22}) becomes negative.

Let us consider the Killing identity for $\phi_b$
\begin{eqnarray}
\nabla_b\nabla^b\phi_a=-R_{a}{}^{b}\phi_b.
\label{nh23}
\end{eqnarray}
Contracting Eq.~(\ref{nh23}) by $\phi^a$ and using Eq.~(\ref{g10})
we get 
\begin{eqnarray}
\nabla_b\nabla^b f^2=\left[ 4\left(\nabla_a
    f\right)\left(\nabla^af\right)-\frac{f^4}{\beta^2}
  \left(\nabla_a\alpha  \right)\left(\nabla^a\alpha\right)-2R_{a
    b}\phi^a\phi^b\right].  
\label{nh24}
\end{eqnarray}
We now project Eq.~(\ref{nh24}) onto $\Sigma$, multiply by
$\Psi_2^2$ and integrate by parts to get 
\begin{eqnarray}
\int_{\Sigma}\beta \left[4f \Psi_2 \left(D_a\Psi_2\right)
  \left(D^af\right)+ 4\Psi_2^2 \left(D_a f\right)
  \left(D^af\right)-\frac{\Psi_2^2f^4}{\beta^2}\left(D_a\alpha
  \right)\left(D^a\alpha\right)-2\Psi^2_2R_{a
    b}\phi^a\phi^b\right]=0. 
\label{nh25}
\end{eqnarray}
Subtracting Eq.~(\ref{nh25}) from Eq.~(\ref{nh22}) we now have
\begin{eqnarray}
\int_{\Sigma}\beta\left[ \left(\beta D_a
    \Psi_1+2\Psi_1D_a\beta\right)^2+ f^2 \left(D_a\Psi_2\right)
  \left(D^a\Psi_2\right)+2\Psi^2_2R_{a b}\phi^a\phi^b 
+m^2\left(\beta^2\Psi_1^2+f^2\Psi_2^2\right)\right]=0.
\label{nh26}
\end{eqnarray}
So the no hair result $\Psi_1=0=\Psi_2$ will follow from
Eq.~(\ref{nh26}) if $R_{ab}\phi^a\phi^b\geq 0$. We have assumed 
that the spacetime satisfies Einstein's equations, so in particular 
\begin{eqnarray}
R_{ab}\phi^a\phi^b=\left(T_{ab}-\frac{1}{2}T
  g_{ab}\right)\phi^a\phi^b+\Lambda f^2. 
\label{nh27a}
\end{eqnarray}
We compute the energy-momentum tensor for the Lagrangian (\ref{nh4}),
\begin{eqnarray}
T_{ab}=F_{ac}F_{b}{}^{c}+m^2A_aA_b +\mathcal{L}g_{ab},
\label{nh27}
\end{eqnarray}
which yields
\begin{eqnarray}
\left(T_{ab}-\frac{1}{2}T
  g_{ab}\right)\phi^a\phi^b=\left(\frac12 b_a^2 + \frac12 f^2e_a^2 +
m^2  f^4 \Psi_2^2 \right),  
\label{nh28}
\end{eqnarray}
where $b_a=F_{ab}\phi^b$ and $e_a$ is the electric field defined in
Eq.~(\ref{nh6}). It is easy to see that $b_a\chi^a=0$, i.e., $b_a$
is spacelike.  The electric field $e^a$ is also spacelike as
mentioned earlier.  So Eq.~(\ref{nh28}) shows that $\left(T_{ab} -
  \frac{1}{2}T g_{ab}\right)\phi^a\phi^b\geq 0$ for the Proca
field.  Putting in all this, we can rewrite Eq.~(\ref{nh26}) as
\begin{eqnarray}
\int_{\Sigma}\beta\left[ \left(\beta D_a \Psi_1+
    2\Psi_1D_a\beta\right)^2+ 
f^2 \left(D_a\Psi_2 \right)\left(D^a\Psi_2\right) 
+m^2\beta^2\Psi_1^2 \right. \qquad \qquad && \nonumber \\ 
 \left. + \left(m^2+2\Lambda\right)f^2\Psi_2^2 +
 2\Psi_2^2\left(\frac12 b_a^2 + \frac12 f^2e_a^2 +
m^2  f^4 \Psi_2^2 \right)
\right] =0,
\label{nh29}
\end{eqnarray}
which gives
$\Psi_1=0=\Psi_2$ over $\Sigma$. Since
$\pounds_{\chi}\Psi_1=0=\pounds_{\chi}\Psi_2$, we have
$\Psi_1=0=\Psi_2$ throughout the manifold. This, combined with the
previous proof $\overline{a}_b=0$, is the desired no hair result
for a de Sitter black hole for the Proca-massive vector field.

Clearly, our proof is also valid for an asymptotically flat
stationary axisymmetric spacetime, $\Lambda=0$. We have only to
replace the outer boundary or the cosmological horizon by a
2-sphere at spacelike infinity with a sufficiently rapid fall off
condition of the fields. Our proof also applies to asymptotically
anti-de Sitter space-time provided we assume $m^2\geq 2|\Lambda|$
in Eq.~(\ref{nh29}) for the asymptotically AdS case. This is not a
strong assumption --- it only means that the Compton wavelength of
the vector field is less than the cosmological length scale or the
AdS radius.

As we have mentioned earlier, the no hair proof fails for $m=0$,
i.e. for the Einstein-Maxwell system, because the local gauge
symmetry implies that $A_a$ is not a physical field, so need not 
be bounded on the horizon.
The Kerr-Newmann-de Sitter spacetime is a black hole solution to
the Einstein-Maxwell equations~\cite{Carter:1968ks}. 
\section{Discussions}
To summarize, we have proven the no hair theorems for scalar and
(Proca) massive vector fields for a stationary axisymmetric de
Sitter black hole spacetime. In comparison to the proof in a static
spacetime, this proof contains some additional constraints such as
the commutativity of the two Killing fields $\xi^a$ and $\phi^a$
and the existence of spacelike 2-submanifolds orthogonal to
them. Also, in order to prove the theorem for the vector field we
had to assume in Eq.~(\ref{nh27a}) that the spacetime satisfies
Einstein's equations. For a static spacetime one need not assume
that (see e.g. \cite{Bhattacharya:2007ap}).


In the static case it is necessary to assume spherical symmetry in
order to prove the no hair theorem for the Abelian Higgs
model~\cite{Lahiri:1993vg, Bhattacharya:2007ap}. In fact if we have
cylindrically symmetric matter distribution we have a cosmic string
piercing the horizons~\cite{Achucarro:1995nu, Bhattacharya:2008fu,
  Bhattacharya:2010ie}. It seems likely that we will have a string
like solution for a rotating axisymmetric de Sitter black hole
as well.


As an aside we note that the no hair results proven are not black
hole uniqueness theorems. It is known that for $\Lambda=0$, the
Kerr spacetime is the only asymptotically flat black hole solution
of the vacuum Einstein equations in 4-dimensions (see
e.g.~\cite{Chandrasekhar:1985kt, Mazur:2000pn} and references
therein). For $\Lambda < 0$ in 2+1 dimensions, a result analogous
to Birkhoff's theorem was proven for the BTZ black
hole~\cite{AyonBeato:2004if}. For $\Lambda > 0$, no proof of
uniqueness of black hole solutions is known~\cite{Boucher:1983cv,
  Robinson}. However, our results reduce the Einstein-scalar (in
convex potential) and Einstein-massive vector (with no gauge
symmetry) systems to vacuum Einstein equations in the presence of a
stationary axisymmetric black hole. So any proof of uniqueness of
the Kerr-de Sitter black hole, if it exists, will apply to these
systems as well.

\section*{Acknowledgment}
SB's work was supported by a fellowship from his institution SNBNCBS.

\vskip 1cm

\end{document}